\documentclass[preprint,12pt,3p]{elsarticle}
\usepackage[utf8]{inputenc} 
\usepackage[T1]{fontenc}    
\usepackage{hyperref}       
\usepackage{url}            
\usepackage{booktabs}       
\usepackage{amsfonts}       
\usepackage{nicefrac}       
\usepackage{microtype}      
\usepackage{lipsum}
\usepackage{amsmath,amssymb,graphicx}
\usepackage{float}
\usepackage{siunitx}
\usepackage{array}
\usepackage{lmodern}
\usepackage{comment}
\usepackage{amssymb}
\usepackage{xcolor}
\usepackage{multicol}
\usepackage{multirow}
\journal{Elsevier}
\begin{document}
\begin{frontmatter}
\title{Nanomechanical behavior of pentagraphyne-based single-layer and nanotubes through reactive classical molecular dynamics}
\author[IFPI,UNICAMP1]{J. M. De Sousa\corref{author}}
\cortext[author]{I am corresponding author}
\ead{josemoreiradesousa@ifpi.edu.br}
\author[UFPI]{W. H. S. Brandão}
\author[CiMa]{W. L. A. P. Silva}
\author[UnB]{\\L. A. Ribeiro Junior}
\ead{ribeirojr@unb.br}
\author[UNICAMP1,UNICAMP2]{D. S. Galv\~ao}
\ead{galvao@ifi.unicamp.br}
\author[UnB2,CiMa]{M. L. Pereira J\'unior}
\ead{marcelo.lopes@unb.br}

\affiliation[IFPI]{organization={Instituto Federal de Educa\c c\~ao, Ci\^encia e Tecnologia do Piau\'i -- IFPI},
            addressline={\\Primavera}, 
            city={São Raimundo Nonato},
            postcode={64770-000}, 
            state={Piauí},
            country={Brazil}}  

\affiliation[UNICAMP1]{organization={Applied Physics Department, ``Gleb Wataghin'' Institute of Physics - IFGW, University of Campinas -- UNICAMP},
            addressline={Rua Sérgio Buarque de Holanda, Cidade Universitária}, 
            city={Campinas},
            postcode={13083-859}, 
            state={São Paulo},
            country={Brazil}}  
\affiliation[UFPI]{organization={Department of Physics, Federal University of Piauí -- UFPI},
            addressline={Ininga}, 
            city={Teresina},
            postcode={64049-550}, 
            state={Piauí},
            country={Brazil}}

\affiliation[CiMa]{organization={University of Brasília, Faculty UnB Planaltina, PPGCIMA},
            city={Brasília},
            postcode={73345-010},
            country={Brazil}} 

\affiliation[UnB]{organization={University of Brasília, Institute of Physics},
            city={Brasília},
            postcode={70910-900},
            country={Brazil}} 

\affiliation[UNICAMP2]{organization={Center for Computing in Engineering and Sciences, University of Campinas},Department and Organization
            addressline={Rua Sérgio Buarque de Holanda, 777 - Cidade Universitária}, 
            city={Campinas},
            postcode={13083-859}, 
            state={São Paulo},
            country={Brazil}}

\affiliation[UnB2]{organization={University of Bras\'{i}lia, Faculty of Technology, Department of Electrical Engineering},
            city={Brasília},
            postcode={70910-900},
            country={Brazil}}

\begin{abstract}
In a recent theoretical study, a new 2D carbon allotrope called pentagraphyne (PG-yne) was proposed. This allotrope is derived from pentagraphene by introducing acetylenic linkages between sp$^3$ and sp$^2$ hybridized carbon atoms. Due to its interesting electronic and structural properties, it is of interest to investigate the mechanical behavior of PG-yne in both monolayer and nanotube topologies. To achieve this, we performed fully atomistic reactive (ReaxFF) molecular dynamics simulations, and our results show that Young's modulus average of PG-yne monolayers is approximately 913 GPa, at room temperature. In comparison, it ranges from 497-789 GPa for the nanotubes studied. Furthermore, we observed that PG-yne monolayers exhibit a direct transition from elastic to complete fracture under critical strain without a plastic regime. In contrast, some PG-yne nanotubes exhibit an extended flat plastic regime before total fracture.
\end{abstract}


\end{frontmatter}
\section{Introduction}
\label{INT}

Carbon-based two-dimensional (2D) materials have gained significant attention due to their unique physical and chemical properties, which are useful in flat electronics \cite{novoselov2004electric}. Among them, graphene stands out as one of the most popular 2D materials \cite{novoselov2004electric}. It consists of a single layer of carbon atoms arranged in a hexagonal lattice, showing exceptional electrical conductivity, mechanical strength, and thermal properties \cite{geim2007rise,balandin2008superior}. These traits combined have made it a promising candidate for a wide range of applications, including electronics \cite{geim2007rise}, energy storage \cite{stoller2008graphene}, and sensors \cite{geim2007rise,neto2009electronic}.

Other 2D carbon allotropes, including $\gamma$-graphyne \cite{hu2022synthesis}, monolayers of biphenylene \cite{fan2021biphenylene}, amorphous carbon \cite{toh2020synthesis}, and fullerene networks \cite{synth_full}, have also been synthesized. Despite these successes, there has been an ongoing effort to develop new materials that can address some of the limitations of graphene, such as its lack of an electronic bandgap, which restricts its use in digital electronic devices.

The discovery of new 2D carbon allotropes has spurred a surge of theoretical investigations \cite{elias2009control}. One such investigation proposed pentagraphene (PG), a material composed entirely of pentagonal carbon rings arranged in a pattern resembling the Cairo pentagonal tiling \cite{zhang2015penta}. This unique material combines high stability, negative Poisson's ratio, and anisotropic conductivity. It also features a semiconducting and indirect band gap of 3.25 eV, and Young's modulus and Poisson's ratio values of approximately 263.8 GPa$\cdot$nm and -0.068, respectively. Although the synthesis of PG has not yet been realized, PG has inspired further studies aimed at developing new materials with similar topology \cite{zhang2015penta}, using fused pentagonal rings that tend to preserve its attractive properties.

A novel 2D carbon allotrope, namely pentagraphyne (PG-yne), was proposed in a theoretical study recently \cite{deb2020pentagraphyne}. PG-yne is more energetically favorable than other graphyne members, including experimentally synthesized graphyne \cite{hu2022synthesis} and graphdiyne \cite{gao2019graphdiyne} monolayers. It is derived from PG by inserting acetylenic linkages between sp$^3$ and sp$^2$ hybridized carbon atoms, in the same way that graphene can generate graphynes. State-of-the-art calculations have shown that PG-yne is dynamically, thermally, and mechanically stable, able to withstand temperatures up to 1000 K. It exhibits intrinsic semiconducting properties, with an electronic bandgap of around 1.0 eV that can be tuned by applying strain. These remarkable properties motivate the comprehensive study of PG-yne's mechanical properties and fracture patterns performed here, which may expand the understanding of the range of applications for structures of similar topology and can stimulate further studies into its synthesis.

In the present study, we have carried out fully atomistic reactive (ReaxFF) classical molecular dynamics (MD) simulations to investigate the mechanical properties and fracture patterns of PG-yne monolayers and nanotubes (PG-yneNTs), as depicted in Figure \ref{fig:memb-nanot}. Our simulations considered a range of tube diameters and temperature values. Young's modulus of PG-yne monolayers was determined to be around 913 GPa at room temperature, whereas the values varied between 497-789 GPa for the nanotubes considered in this study. When a critical strain was reached, both PG-yne and $(n,0)$PG-yneNTs underwent complete fracture without exhibiting a plastic deformation stage. Conversely, $(n,n)$PG-yneNTs exhibited a flat plastic region between the elastic and completely fractured regimes. 

\section{Methodology}

\begin{figure}[htb!]
    \centering
    \includegraphics[width=\linewidth]{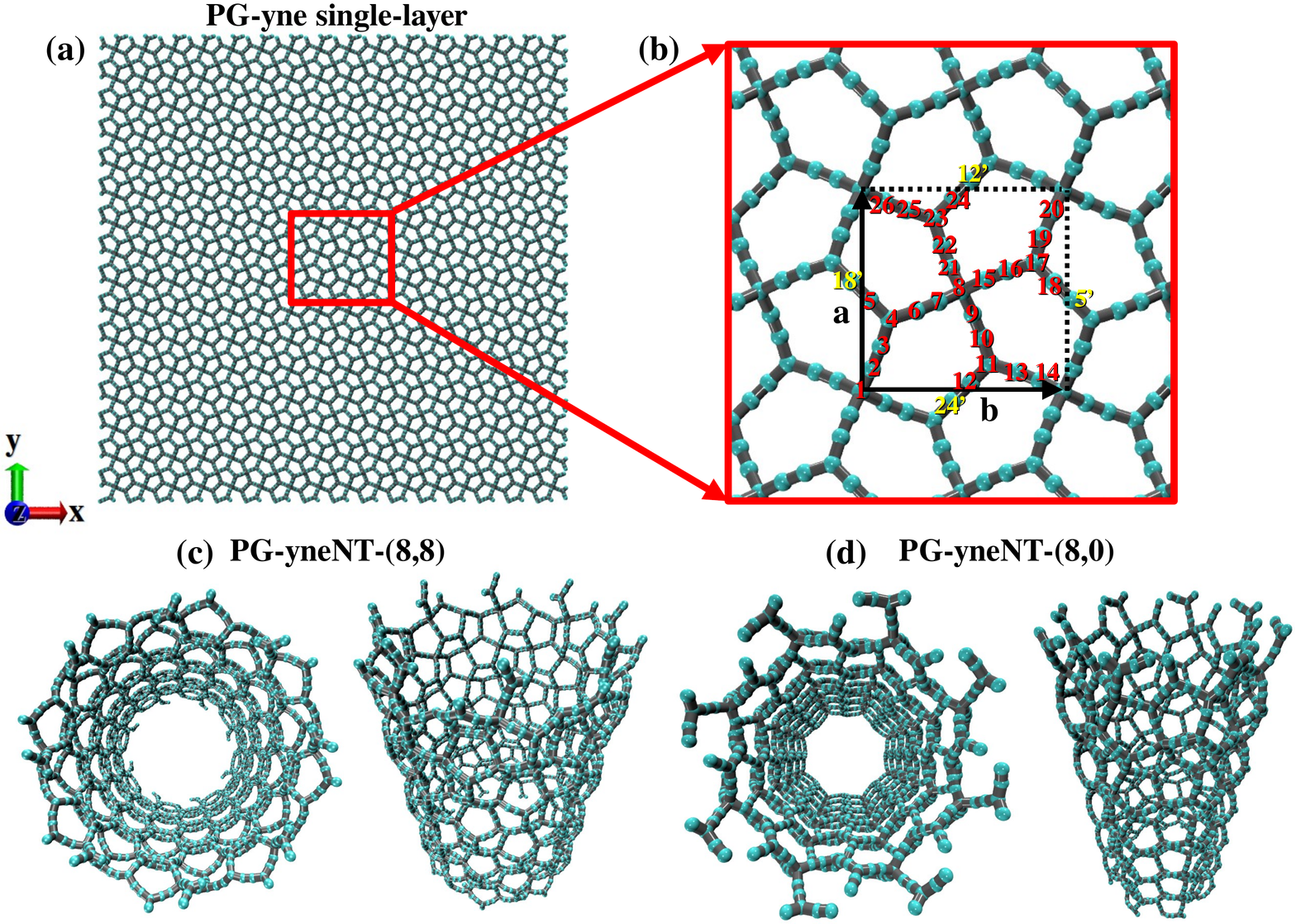}
    \caption{Schematic representation of (a) PG-yne monolayer, (b) PG-yne unit cell highlighted by a black square with 26 atoms (red labels), where $\textbf{a} = \textbf{b} = 9.85$\r{A}, (c) $(8,8)$-PG-yneNT, and (d) $(12,0)$-PG-yneNT. See Tab \ref{tab:param-pg-yne} for structural information.}
    \label{fig:memb-nanot}
\end{figure}

To investigate the mechanical properties and fracture patterns of PG-yne monolayers and PG-yneNTs, we performed fully atomistic reactive molecular dynamics (MD) simulations using the LAMMPS \cite{plimpton1995fast} code with the ReaxFF \cite{van2001reaxff} potential. ReaxFF is a widely used interatomic force field for studying the mechanical properties of nanostructures since it can handle bond formation and breaking at the atomic level \cite{mueller2010development}. In our model of the PG-yne supercell  and for the nanotubes, we simulate the chirality types $(n,n)$ and $(n,0)$, shown in Figure \ref{fig:memb-nanot}. Structural details of the PG-yne and PG-yneNTs are presented in Table \ref{tab:param-pg-yne}.

When creating the PG-yneNTs models computationally, we define the chiral vector $\textbf{C}_h$ as $n\textbf{a}+m\textbf{b}$, where $\textbf{a}$ and $\textbf{b}$ are the lattice vectors and $n$ and $m$ are integers that determine the chirality. The diameter of the nanotube is calculated using $d_t=|\textbf{C}_h|/\pi$. The translational vector $\textbf{T}$, perpendicular to $\textbf{C}_h$, is determined by $\textbf{T}=t_1\textbf{a}+t_2\textbf{b}$, where $t_1$ and $t_2$ are integers obtained through the inner product $\textbf{C}_h\cdot\textbf{T}=0$. The length of the nanotube is $L=|\textbf{T}|$, and the chiral and translational vectors define the nanotube unit cell. Additionally, since PG-yne's unit cell is a square ($\textbf{a}=\textbf{b}$), configurations $(m,0)$ and $(0,m)$ for PG-yneNTs are the same, except for a rotation.

\begin{table}[htb!]
    \centering
    \begin{tabular}{|c|c|c|c|c|}
    \hline
        \multicolumn{5}{|c|}{Unit cell} \\ \hline
        \multicolumn{2}{|c|}{Angles} & $\measuredangle$ ($^\circ$) & Bonds & $d_{\mathrm{C-C}}$ (\AA)\\ \hline
        \multicolumn{2}{|c|}{C$_1$-C$_2$-C$_3$} & 165.19 & \multirow{2}{*}{C$_1$-C$_2$} & \multirow{2}{*}{1.48}\\ \cline{1-3}
        \multicolumn{2}{|c|}{C$_2$-C$_3$-C$_4$} & 171.69 & & \\ \hline
        \multicolumn{2}{|c|}{C$_3$-C$_4$-C$_5$} & 118.92 & \multirow{2}{*}{C$_2$-C$_3$} & \multirow{2}{*}{1.22} \\ \cline{1-3}
        \multicolumn{2}{|c|}{C$_3$-C$_4$-C$_6$} & 120.70 & & \\ \hline
        \multicolumn{2}{|c|}{C$_4$-C$_5$-C$'_{18}$} & 169.22 & \multirow{2}{*}{C$_3$-C$_4$} & \multirow{2}{*}{1.42}\\ \cline{1-3}
        \multicolumn{2}{|c|}{C$_4$-C$_6$-C$_7$} & 171.69 & & \\ \hline
        \multicolumn{2}{|c|}{C$_7$-C$_8$-C$_9$} & 106.88 & C$_4$-C$_5$ & 1.36 \\ \hline
        \multicolumn{2}{|c|}{C$_7$-C$_8$-C$_{15}$} & 115.00 & C$_5$-C$'_{18}$ & 1.25\\ \hline
        \multicolumn{5}{|c|}{Super cell PG-yne} \\ \hline
        \multicolumn{2}{|c|}{$n_a\times n_b$} & Number of Atoms & Length $x$ & Length $y$ \\ \hline
        \multicolumn{2}{|c|}{$16\times16$} & 6656 & 157.16 & 157.16 \\ \hline
        \multicolumn{5}{|c|}{PG-yneNTs} \\ \hline
         Chirality & $n$ & Number of Atoms & Diameter (\r{A}) & Length (\r{A}) \\ \hline
        \multirow{3}{*}{ $(n,0)$}& 4     & 624		& 12.54		& 59.10 \\ \cline{2-5}
        & 8 	& 1248		& 25.08		& 59.10 \\ \cline{2-5}
        & 12    & 1872		& 37.62		& 59.10 \\ \hline
        \multirow{3}{*}{ $(n,n)$} &  3	&  624		& 13.30		& 55.72 \\ \cline{2-5}
        & 6	& 1248		& 26.60		& 55.72 \\ \cline{2-5}
        & 8	& 1664		& 35.47		& 55.72 \\ \hline
    \end{tabular}
    \caption{Structural parameters of the unit cell and supercell PG-yne and PG-yneNTs. In the unit cell (see Fig. \ref{fig:memb-nanot}), the distances C-C is represented by $d_{\mathrm{C-C}}$. The integers $n_a$ and $n_b$ represent repetitions along vectors \textbf{a} and \textbf{b}, respectively.}
    \label{tab:param-pg-yne}
\end{table}

In our computational approach, we first minimized the energy of the PG-yne and PG-yneNTs systems. Subsequently, we coupled them to a thermostat chain for thermodynamic equilibrium. We performed constant NPT ensemble integration at null pressure and at 300 K to ensure no remaining stress, using a Nose/Hoover \cite{evans1985nose} pressure barostat for 25 ps. Following this, we coupled all the systems in the canonical NVT ensemble for an additional 25 ps. This enabled the generation of sampled positions and velocities for a range of temperatures from 10 K up to 1200 K.

To investigate the overall mechanical behavior of PG-yne and PG-yneNTs, we conducted tensile tests by stretching the systems until rupture. The tests were performed using a constant engineering tensile strain rate of $10^{-6}/$fs within an NVT ensemble. Stress was applied to PG-yne along the x-direction (layers) and to PG-yneNTs along the z-direction (tubes), while the other directions were allowed to relax freely. We analyzed the stress-strain curves to extract the elastic properties of each structure at different temperatures ranging from 10 K up to 1200 K. We also described the bond breaks and defined the fracture patterns of PG-yne and PG-yneNTs structures by analyzing the MD snapshots. To visualize the MD snapshots, we used VMD \cite{humphrey1996vmd} software and the nanotube structures were generated using a Fortran code\cite{backus1964fortran}. 

\section{Results}

\begin{figure}[htb!]
    \centering
    \includegraphics[scale=0.6]{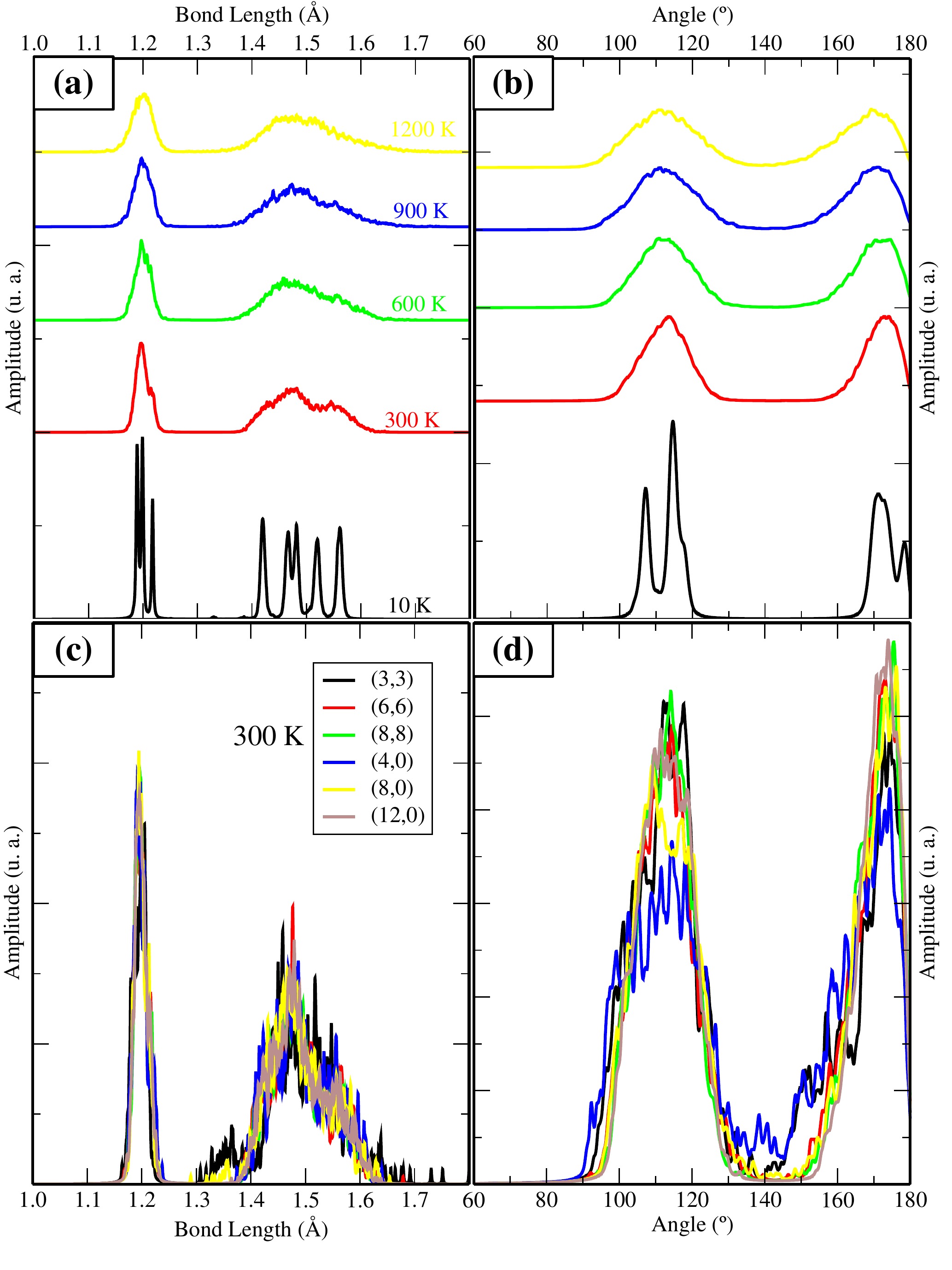}
    \caption{PG-yne (a) bond and (b) angle distributions (in arbitrary units), as a function of the temperature after lattice thermalization. Black, red, green, blue, and yellow lines refer to the $10$ , $300$ , $600$ , $900$ , and $1200$ K cases, respectively.
    }
    \label{fig:dist-bond-ang}
\end{figure}

We start the discussion by analyzing the bond and angle distributions of PG-yne and PG-yneNTs after lattice thermalization at different temperatures. The corresponding profiles are shown in Figure \ref{fig:dist-bond-ang}, where black, red, green, blue, and yellow lines represent the 10, 300, 600, 900, and 1200 K cases, respectively. In Figure \ref{fig:dist-bond-ang}(a), the bond distribution profile displays a prominent peak centered at 1.2 \r{A} (sp$^{2}$-hybridized C$_2$-C$_3$ and C$_5$-C$'_{18}$  bonds), and four similar peaks within the range of 1.4-1.6 \r{A} (sp$^{2}$-hybridized C$_1$-C$_2$, C$_3$-C$_4$ and C$_4$-C$_5$ bonds) at 10 K. On the other hand, in Figure \ref{fig:dist-bond-ang}(b), the angle distribution profile exhibits two peaks between 100$^\circ$-120$^\circ$ (C$_3$-C$_4$-C$_5$, C$_3$-C$_4$-C$_6$, C$_7$-C$_8$-C$_9$ and C$_7$-C$_8$-C$_{15}$ angles) and 160$^\circ$-180$^\circ$ (C$_1$-C$_2$-C$_3$, C$_2$-C$_3$-C$_4$, C$_4$-C$_5$-C$'_{18}$ and C$_4$-C$_6$-C$_7$ angles) at 10 K. In turn, the two types of nanotubes, $(n,n)$ and $(n,0)$, exhibit practically the same curves, overlapping each other, for the distributions of bonds and angles (Fig. \ref{fig:dist-bond-ang}(c)-(d)). Furthermore, these distributions show peaks around intervals practically equal to those seen for membranes at 300 K.

The C$-$C bond-length values increase with temperature due to larger amplitude thermal vibrations of the carbon atoms. In this sense, the bond distribution profile changes, with the most prominent peak holding at the same position and the four peaks merging into a single one centered at 1.5 \r{A} for temperatures between 100-1200 K. These results are consistent with the original work and indicate the structural stability of PG-yne with the coexistence of sp$^2$ and sp$^3$-like bonds. PG-yne retains its planar morphology, with two well-defined angle distributions centered at 110$^\circ$ and 170$^\circ$ for temperatures between 100-1200 K, as shown in Figure \ref{fig:dist-bond-ang}(b).  

\begin{figure}[htb!]
    \centering
    \includegraphics[width=\linewidth]{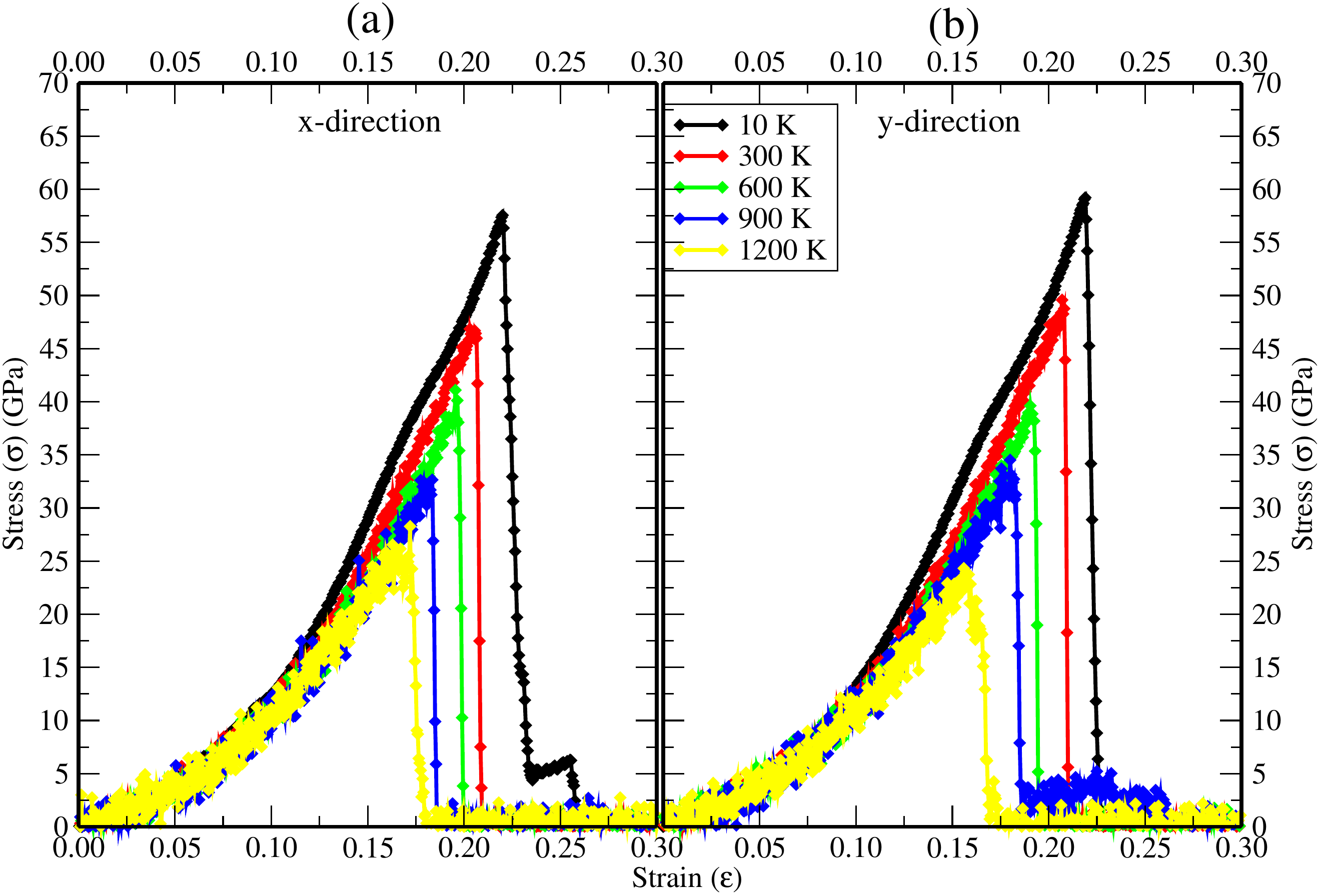}
    \caption{PG-yne stress-strain curves for different temperatures.}
    \label{fig:ss-meb}
\end{figure}

Figure \ref{fig:ss-meb} displays stress-strain curves for PG-yne at temperatures ranging from 10 up to 1200 K. The left and right panels show the x- and y-direction stretching profiles. Black, red, green, blue, and yellow lines indicate the 10, 300, 600, 900, and 1200 K cases. These curves indicate that the PG-yne monolayers transition directly from elastic to completely fractured regimes at a critical strain ($\epsilon_C$), which is also known as the fracture strain. The stress-strain curves show a similar trend for tensile loading applied along the x- or y-direction, in line with the topology of PG-yne. Higher temperatures result in lower values of $\sigma_C$, the critical stress at the first fracture, indicating that PG-yne becomes easier to fracture as temperature increases. This can be attributed to the greater amplitude and probability of bond breaking due to more pronounced thermal vibrations of carbon atoms in the nanostructure at higher temperatures.

\begin{figure}[htb!]
    \centering
    \includegraphics[width=\linewidth]{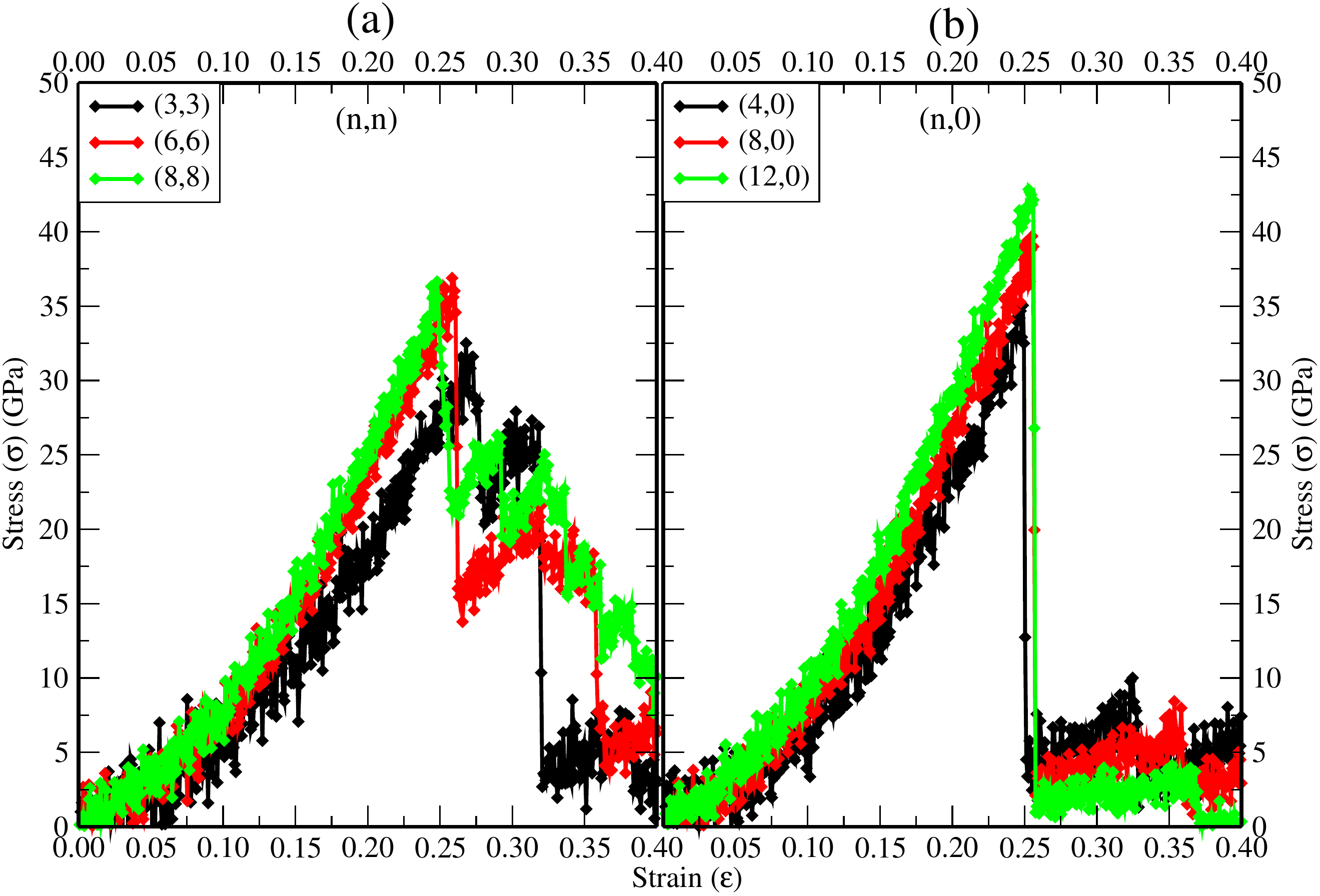}
    \caption{PG-yneNTs stress-strain curves at 300 K.}
    \label{fig:ss-nanot}
\end{figure}

PG-yneNTs can exhibit a plastic stage before fracturing, illustrated in Figure \ref{fig:ss-nanot}. Their topology significantly influences the mechanical properties. The left and right panels of the figure show the stress-strain interplay for the $(n,n)$ and $(n,0)$ cases, respectively, when subjected to stretching along their longitudinal (z-direction). In the left panel, the (3,3), (6,6), and (8,8) cases are represented by black, red, and green lines, respectively. In contrast, the right panel shows the (4,0), (8,0), and (12,0) cases at 300 K. The distinct curve profiles indicate that PG-yneNTs can withstand more strain than PG-yne layers before fracturing. This behavior can be attributed to the tubular topology of PG-yneNTs, which provides many pathways for stress dissipation, thus preventing early and brittle fractures compared to PG-yne monolayers.

The mechanical properties of $(n,n)$-type and $(n,0)$-type PG-yneNTs differ due to the different arrangements of carbon atoms, leading to variations in bond angles and bond lengths. The $(n,0)$-type PG-yneNTs possess sp$^3$-hybridized C-C bonds aligned along the direction of the applied strain, resulting in a higher resistance to deformation than the $(n,n)$-type PG-yneNTs. The $(n,n)$-type PG-yneNTs contain sp$^2$-like C-C bonds aligned with the tube axis, which resist deformation and fracture by forming linear atomic chains (LACs) at about 25\% within the plastic stage. However, the $(n,0)$-type PG-yneNTs have these bonds oriented at an angle to the tube axis, with average fracture strain also close to 25\%, but with slightly higher $\sigma_C$ (see Table \ref{tab:elastic-val}). The fracture strain in $(n,0)$-type PG-yneNTs is not influenced by the tube diameter, and their stress-strain interplay is similar to the ones for PG-yne monolayers. Figure \ref{fig:memb-nanot}(c) and (d) provide visual representations of the bond configurations in PG-yneNTs along their longitudinal axis.

The PG-yneNTs have been found to withstand higher strain values before breaking compared to PG-yne monolayers, but they can withstand lower critical stress values, as shown in Table \ref{tab:elastic-val}. Our simulations for all temperatures indicate that the average of $Y_{Mod}$ for PG-yne monolayers is approximately 823.92 GPa for direction $x$ and 897.81 GPa for direction $y$. At the same time, PG-yneNTs have averages of 731.69 GPa and 642.72 GPa for types $(n,n)$ and $(n,0)$, respectively. Moreover, smaller-diameter nanotubes tend to be stiffer and have higher Young's modulus ($Y_M$) but smaller critical strains. The $Y_M$ value for PG-yne monolayers is approximately 913 GPa for room temperature, while for PG-yneNTs, it can range from 479-789 GPa. It is worth noting that these values are comparable to those reported for other carbon-based 2D materials \cite{brandao2021mechanical,junior2021reactive,junior2020elastic} and are lower than those for graphene \cite{jiang2009young}, mainly because of the porous structure of PG-yne.
\begin{table}[htb!]
    \centering
    \begin{tabular}{|c|c|c|c|c|c|}
    \hline
    \multicolumn{6}{|c|}{PG-yne} \\ \hline
     DIR. &   TEMP. (K) & $Y_{Mod}$ (GPa) & US (GPa) & $\sigma_C$ (GPa) & $\varepsilon_C$ \\ \hline
      \multirow{5}{*}{$x$}
        & 10   & 936.59 $\pm$ 1.04 & 57.55 $\pm$ 0.18& 56.35 $\pm$ 0.18 & 0.22  \\ \cline{2-6}
        & 300  & 834.74 $\pm$ 3.26 & 46.85 $\pm$ 0.30& 41.71 $\pm$ 0.70 & 0.21  \\ \cline{2-6}
        & 600  & 756.06 $\pm$ 4.39 & 41.09 $\pm$ 0.43& 35.41 $\pm$ 0.73 & 0.20  \\ \cline{2-6}
        & 900  & 774.54 $\pm$ 5.32 & 32.65 $\pm$ 0.29& 32.65 $\pm$ 0.29 & 0.18  \\ \cline{2-6}
        & 1200 & 817.70 $\pm$ 6.91 & 28.27 $\pm$ 0.63& 28.27 $\pm$ 0.63 & 0.17  \\ \hline
       \multirow{5}{*}{$y$} 
        & 10   & 968.10 $\pm$ 1.13 & 59.21 $\pm$ 0.17& 59.21 $\pm$ 0.17 & 0.22  \\ \cline{2-6}
        & 300  & 990.43 $\pm$ 2.94 & 49.58 $\pm$ 0.34& 43.92 $\pm$ 0.74 & 0.21  \\ \cline{2-6}
        & 600  & 819.17 $\pm$ 5.02 & 39.63 $\pm$ 0.21& 18.97 $\pm$ 2.94 & 0.19  \\ \cline{2-6}
        & 900  & 792.01 $\pm$ 6.05 & 34.52 $\pm$ 0.45& 17.02 $\pm$ 2.31 & 0.18  \\ \cline{2-6}
        & 1200 & 919.34 $\pm$ 6.78 & 24.32 $\pm$ 0.28& 12.94 $\pm$ 0.83 & 0.17  \\ \hline
        \multicolumn{6}{|c|}{PG-yneNTs (300 K)} \\ \hline
        TYPE &   $(n,m)$ & $Y_{Mod}$ (GPa) & US (GPa) & $\sigma_C$ (GPa) & $\varepsilon_C$ \\ \hline
        \multirow{3}{*}{$(n,n)$}
        & (3,3)   & 782.92 $\pm$ 14.47 & 32.52 $\pm$ 0.69 & 31.59 $\pm$ 0.74 & 0.27  \\ \cline{2-6}
        & (4,4)   & 623.41 $\pm$ 11.59 & 36.89 $\pm$ 0.49 & 34.57 $\pm$ 0.29 & 0.26  \\ \cline{2-6}
        & (8,8)   & 788.76 $\pm$  7.77 & 36.65 $\pm$ 0.30 & 26.16 $\pm$ 0.23 & 0.29  \\ \hline \hline
        \multirow{3}{*}{$(n,0)$}
        & (4,0)  & 496.94 $\pm$ 14.18 & 35.05 $\pm$ 0.27 & 32.50 $\pm$ 0.38 & 0.25  \\ \cline{2-6}
        & (8,0)  & 689.26 $\pm$ 10.19 & 39.70 $\pm$ 0.52 & 39.01 $\pm$ 0.50 & 0.26  \\ \cline{2-6}
        & (12,0) & 741.96 $\pm$  7.23 & 42.85 $\pm$ 0.30 & 42.14 $\pm$ 0.20 & 0.26  \\ \hline
    \end{tabular}
    \caption{Elastic properties for PG-yne and PG-yneNTs estimated from the stress-strain curves presented in Figures \ref{fig:ss-meb} and Figure \ref{fig:ss-nanot}.}
    \label{tab:elastic-val}
\end{table}

\begin{figure}[htb!]
    \centering
    \includegraphics[scale=0.5]{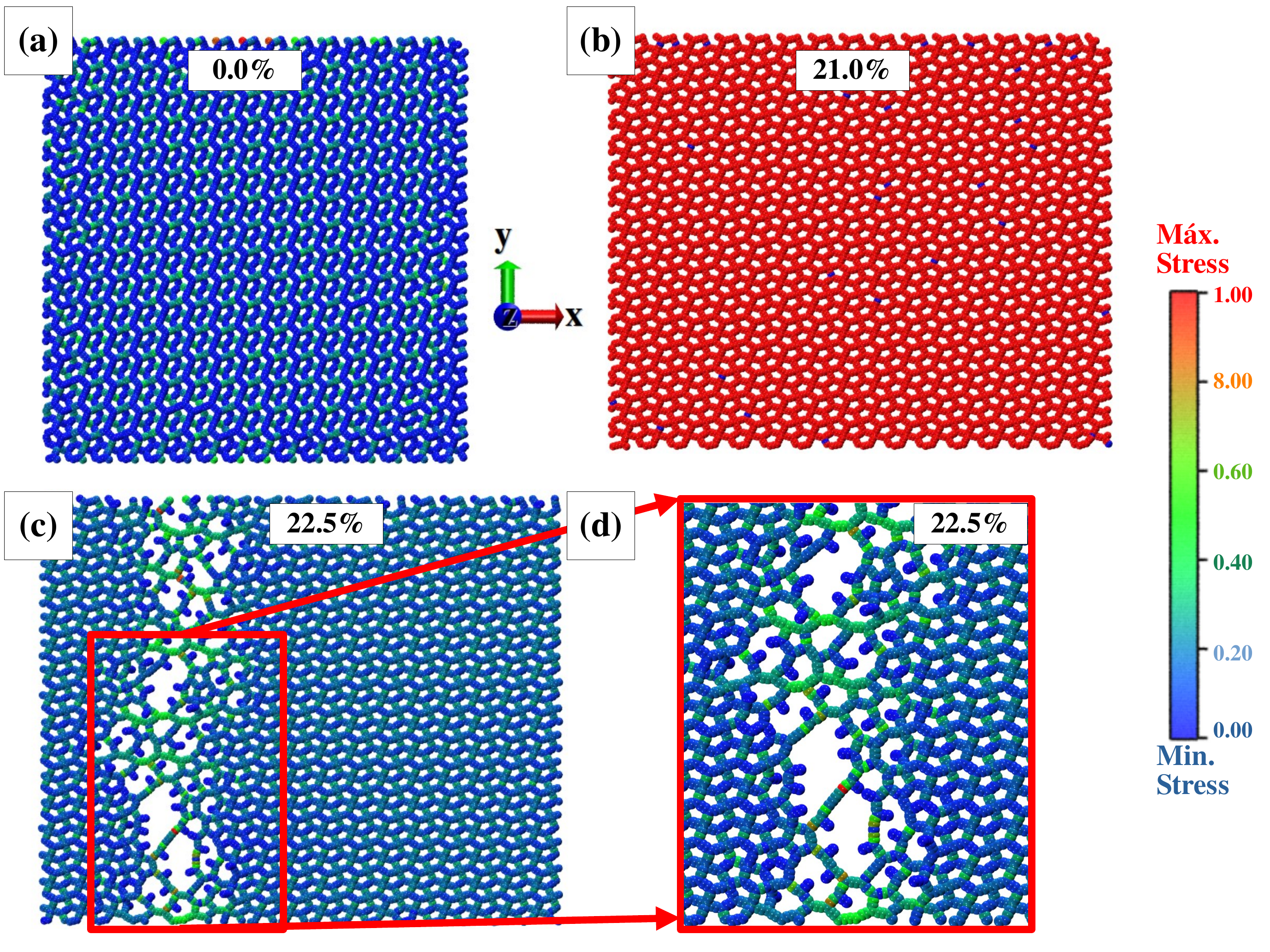}
    \caption{Representative MD snapshots of a PG-yne monolayer subjected to the uniaxial strain in the $x$-direction at $10$ K. In panel (a), the strain is at $0$\%, while in panel (b), it's at $21.0$\%. (c) depicts the fracture at $22.5$\% strain, and The formation of short carbon chains is seen in part (d) for 22.5\%. In the color scheme used for these snapshots, red indicates high-stress accumulation, while blue indicates low-stress accumulation. 
    }
    \label{fig:evo-memb-x}
\end{figure}

\begin{figure}[htb!]
    \centering
    \includegraphics[scale=0.5]{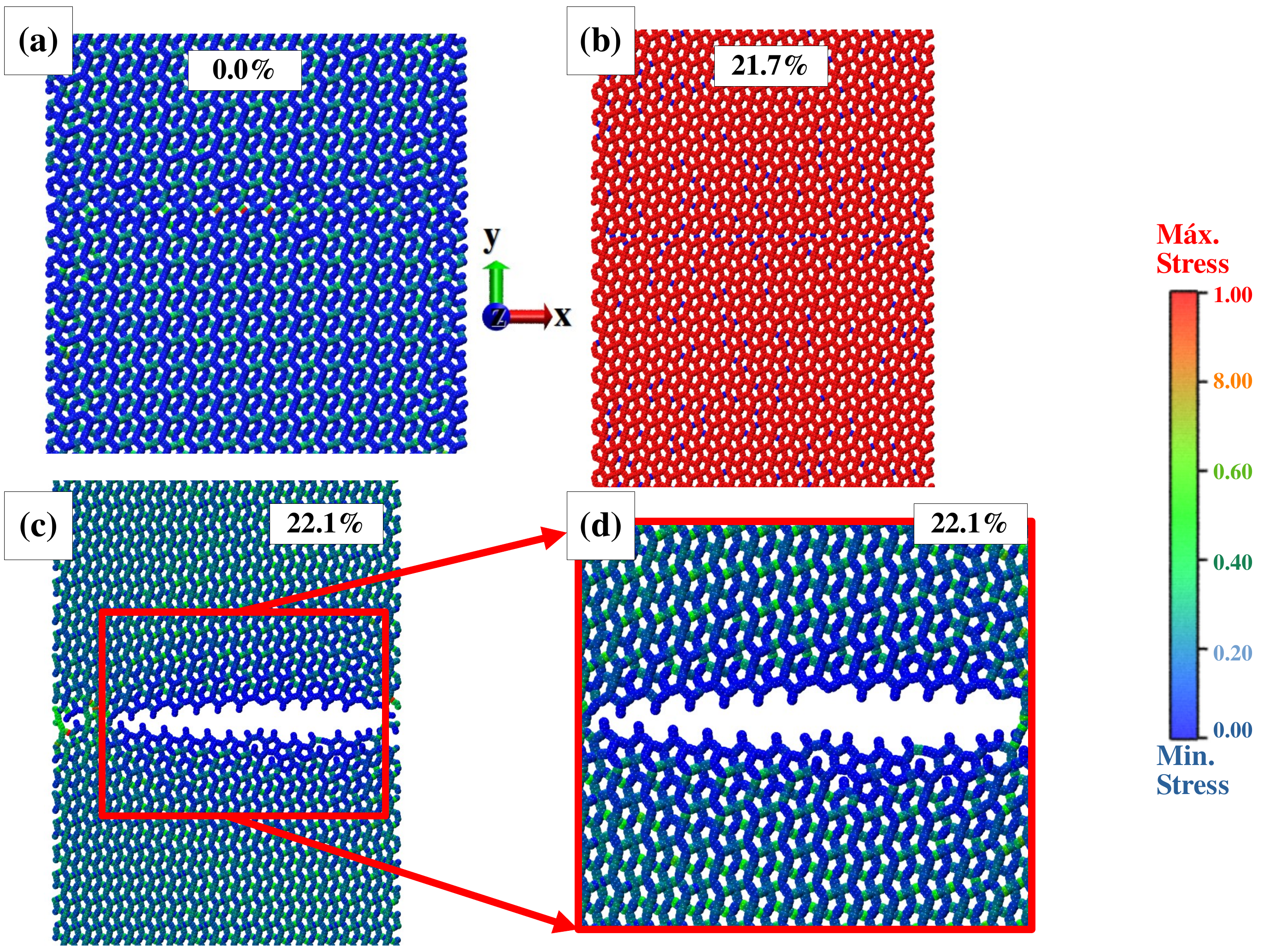}
    \caption{Representative MD snapshots of a PG-yne monolayer subjected to the uniaxial strain along the $y$-direction at $10$ K. In panel (a), the strain is at $0$\%, while in panel (b), it is at $21.7$\%. (c) depicts fracture at $22.1$\% strain, and panel (d) shows the patterns of fracture with no breaking of chemical bonds and breaking of chemical bonds to $22.1$\% of strain. 
    }
    \label{fig:evo-memb-y}
\end{figure}

We also analyzed the fracture patterns of PG-ynes using MD simulations. Figures \ref{fig:evo-memb-x} and \ref{fig:evo-memb-y} illustrate typical MD snapshots of the fracture process for PG-yne under x- and y-directional stress at 10 K, respectively. The color-coding scheme in these figures shows the von Mises (VM) stress per-atom values \cite{mises_NGWZGM}, where red and blue colors indicate high- and low-stress accumulation, respectively. These VM values help identify the points or regions where the fracture starts or propagates. More details on VM calculations can be found in references \cite{junior2022torsional}.

Figure \ref{fig:evo-memb-x}(a) shows the unstressed PG-yne monolayer, dominated by a blue color indicating low stress. At 21.00\% of strain (Figure \ref{fig:evo-memb-x}(b)), the monolayer exhibits a moderate and uniform stress accumulation, represented by a green color, without any visible signs of fracturing. Once the strain surpasses a critical threshold, the fracture occurs, resulting in the breaking of carbon-carbon bonds in the sheet and rapid propagation of cracks that separate the sheet into smaller fragments, as shown in Figure \ref{fig:evo-memb-x}(c) at 22.5\% of strain. In carbon-based 2D materials, it is typical for cracks to propagate along the opposite direction of the applied strain \cite{junior2023irida,pereira2022mechanical,pereira2020elastic,junior2020thermomechanical}. The formation of LACs, shown in Figure \ref{fig:evo-memb-x}(c), is insufficient to establish a distinct plastic stage in the fracture process of PG-yne sheets. Figures \ref{fig:evo-memb-x}(d) and \ref{fig:evo-memb-x}(e) display the lattice arrangement in the region where the first bond breaking occurred for the critical fracture strain (22.00\%) and a higher strain value (22.01\%), respectively. These snapshots reveal that the bonds oriented close to the strain direction are the first ones to break. 

In Figure \ref{fig:evo-memb-y}, the fracture patterns of PG-yne under y-directional stress at 10 K are shown, and they follow the same trend observed along the x-directional stress. The unstressed PG-yne monolayer is presented in Figure \ref{fig:evo-memb-y}(a). At 21.01\% of strain (Figure \ref{fig:evo-memb-y}(b)), the monolayer experiences a moderate and uniform stress accumulation, indicated by green colors, without any visible signs of fracturing. However, at 22.15\% of strain, fast crack propagation leads to fracture, as shown in Figure \ref{fig:evo-memb-y}(c), without any LAC formation. The lattice arrangement in the region where the first bond breaking occurred for the critical fracture strain (21.85\%) and a higher strain value (21.95\%) is presented in Figures \ref{fig:evo-memb-y}(d) and \ref{fig:evo-memb-y}(e), respectively, revealing that the bonds oriented close to the strain direction are the first ones to break.

\begin{figure}[htb!]
    \centering
    \includegraphics[scale=0.5]{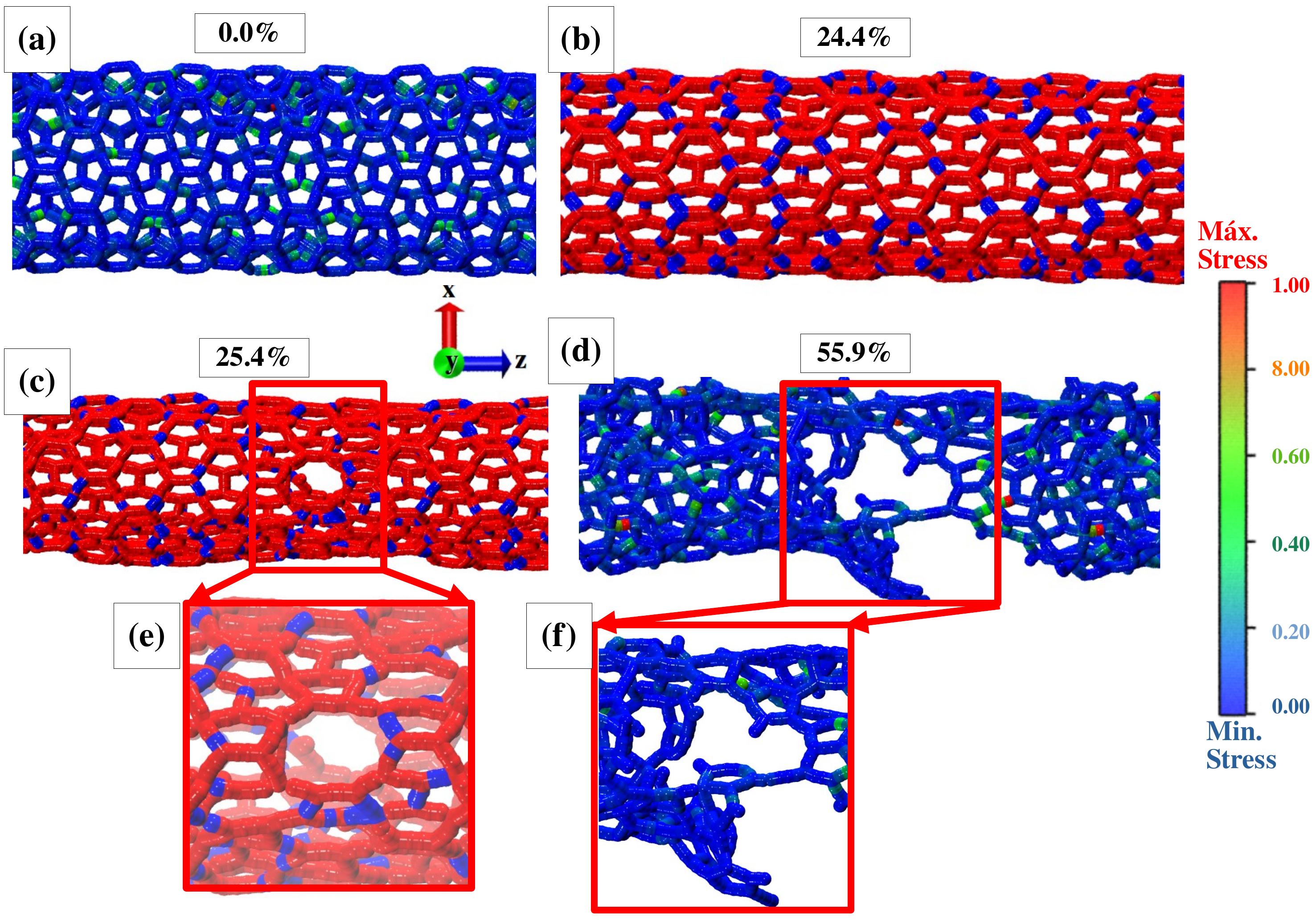}
    \caption{Representative MD snapshots of an (8,8)PG-yneNT subjected to the longitudinal strain in the $z$-direction at $10$ K. In panel (a), the strain is at $0$\%, while in panel (b), it's at $24.4$\%. Fracture starts at $25.4$\% strain (c). Part (d) indicates another stage of fracture at 55.9\% of strain. A zoom of the structure is presented in parts (e) and (f). 
    }
    \label{fig:evo-nanot-n-n}
\end{figure}

\begin{figure}[htb!]
    \centering
    \includegraphics[scale=0.5]{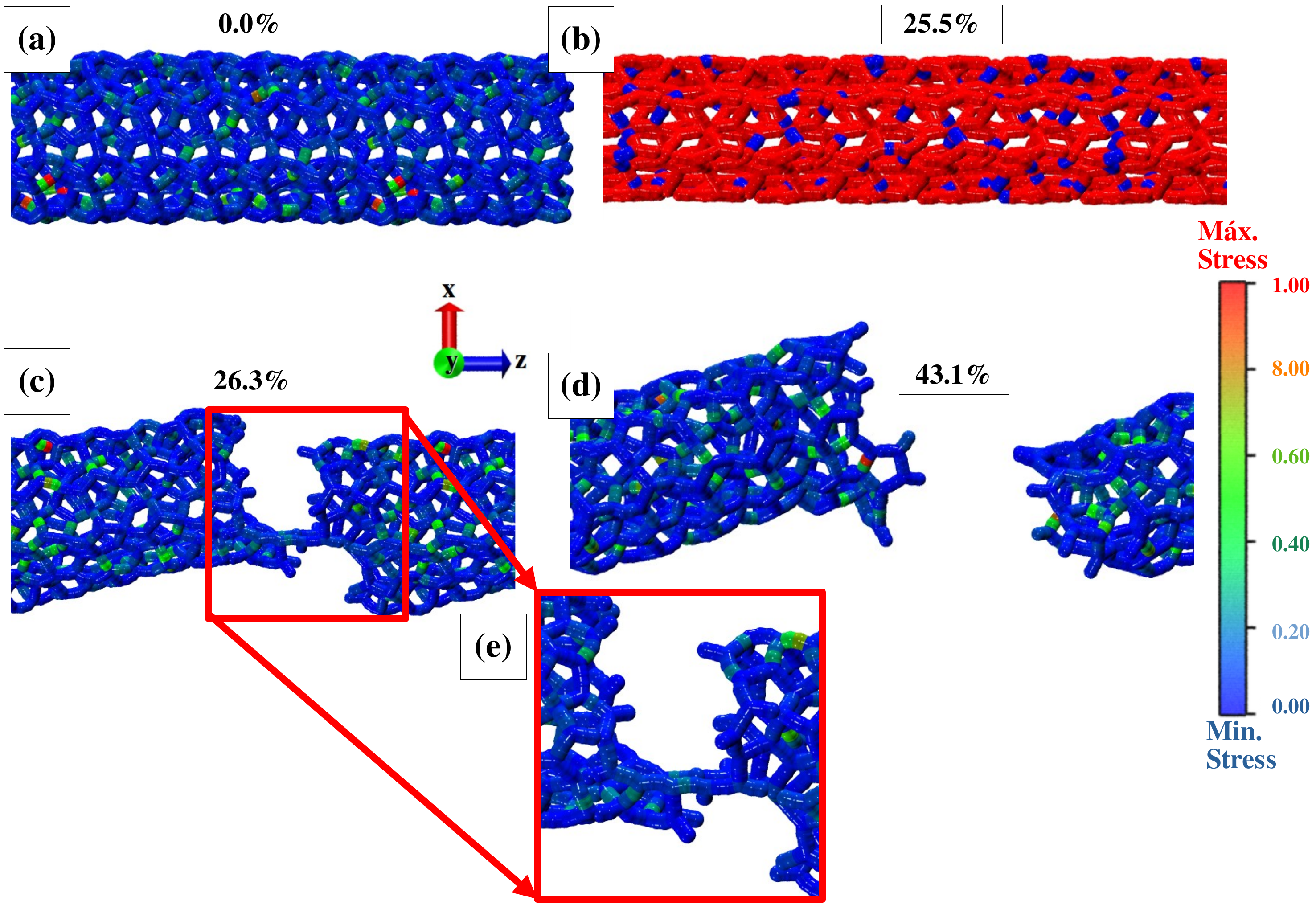}
    \caption{Representative MD snapshots of an (8,0)PG-yneNT subjected to a longitudinal strain along the $z$-direction at $10$ K. In panel (a), the strain is at $0$\%, while in panel (b), it is at $25.5$\%. The fracture starts at 26.3\% (c). Part (d) depicts a complete fracture at $43.1$\% strain, and part (e) shows a zoom of the structure at 26.3\% of strain. 
    }
    \label{fig:evo-nanot-n-0}
\end{figure}

Finally, we present some MD snapshots for the fracture process of the PG-ynes studied here. Again, we use a color scheme representing the VM stress per-atom values to help visualization of the fracture process. In this way, Figures \ref{fig:evo-nanot-n-n} and \ref{fig:evo-nanot-n-0} show the representative cases (8,8)PG-yneNT and (8,0)PG-yneNT, respectively. As a general trend, the fracture patterns obtained in the simulations depend on the chirality of the tubes, as expected.

Figures \ref{fig:evo-nanot-n-n}(a) and \ref{fig:evo-nanot-n-n}(b) show the unstressed and partially stressed (8,8)PG-yneNT at 24.46\% strain, respectively. The stress is primarily accumulated in the sp$^2$-like C-C bonds, which are also the first to break. Long LACs form as a result, as seen in Figure \ref{fig:evo-nanot-n-n}(c) at 50.59\% of strain. The stretching dynamics cause the tube wall to almost collapse, as illustrated in Figure \ref{fig:evo-nanot-n-n}(d). The bonds in $(n,n)$PG-yne are oriented at an angle to the tube axis, leading to uneven load distribution and more complex tube fracture patterns with multiple cracks and LACs. Fragmentation occurs due to the formation of new free edges, which can nucleate new cracks and cause the structure to break into smaller pieces. This mechanism defines the plastic stage seen in the stress-strain curves of $(n,n)$PG-yneNTs (refer to Figure \ref{fig:evo-nanot-n-n}). Snapshots of the lattice arrangement in the region where the first bond breaking occurred for the critical fracture strain (19.26\%) and a higher strain value (31.25\%) are presented in Figures \ref{fig:evo-memb-x}(d) and \ref{fig:evo-memb-x}(e), respectively. These snapshots demonstrate that the bonds oriented at a certain angle to the strain direction are the first ones to break.  

The fracture pattern of $(n,0)$PG-yneNTs is shown in Figure \ref{fig:evo-nanot-n-0}. Figure \ref{fig:evo-nanot-n-0}(a) and \ref{fig:evo-nanot-n-n}(b) display the unstressed and partially stressed (8,0)PG-yneNT at 22.30\% strain, respectively. The stress tends to be distributed almost equally throughout the structure, and no LACs are formed (refer to Figure \ref{fig:evo-nanot-n-n}(c) at 30.29\% of strain). The tube wall does not collapse during the stretching dynamics, as shown in Figure \ref{fig:evo-nanot-n-n}(d). Snapshots of the lattice rearrangement in the region where the first bond breaking occurred for the critical fracture strain (25.49\%) and a higher strain value (25.65\%) are presented in Figures \ref{fig:evo-nanot-n-0}(e) and \ref{fig:evo-nanot-n-0}(f), respectively. These snapshots reveal that the bonds perpendicular to the strain direction are the first ones to break (Figure \ref{fig:evo-nanot-n-0}(e)), followed by the bonds aligned to the strain direction that break almost simultaneously, as illustrated in Figure \ref{fig:evo-nanot-n-0}(f). 

\section{Conclusions}

In summary, we have used fully atomistic reactive (ReaxFF) MD simulations to investigate the mechanical properties and fracture patterns of PG-yne and PG-yneNTs systems. Our simulations encompassed a range of tube diameters and temperature values. As a general trend, our findings revealed that topology significantly influences the mechanical properties of PG-yne 1D (nanotubes) and 2D (monolayers) structures. PG-yne monolayers transition directly from elastic to completely fractured regimes at a critical strain. The stress-strain interplay shows a similar trend for tensile loading applied along the x- or y-direction of the PG-yne monolayer, in line with its topology. Higher temperatures result in lower critical strain values, indicating that PG-yne becomes easier to fracture as temperature increases. 

PG-yneNTs, in turn, can exhibit a plastic stage before fracturing. Their topology significantly influences the mechanical properties. PG-yneNTs can withstand more strain than PG-yne layers before fracturing. This behavior can be attributed to the tubular topology of PG-yneNTs, which provides many pathways for stress dissipation, thus preventing early and brittle fractures compared to PG-yne monolayers. The mechanical properties of $(n,n)$-type and $(n,0)$-type PG-yneNTs differ due to the different arrangements of carbon atoms, leading to variations in bond angles and bond lengths. 

The $(n,n)$-type PG-yneNTs contain sp$^2$-like C-C bonds aligned with the tube axis, which resist deformation and fracture by forming linear atomic chains at about 45\%, within the plastic stage. However, the $(n,0)$-type PG-yneNTs have these bonds oriented at an angle to the tube axis, making them easier to fracture at about 25\%, but with slightly higher critical strain. The fracture strain in $(n,0)$-type PG-yneNTs is not influenced by the tube diameter, and their stress-strain interplay is similar to the ones for PG-yne monolayers. 

Our simulations indicate that the critical strain for PG-yne monolayers is approximately 198.33 GPa, while for PG-yneNTs, it ranges from 108.81 GPa up to 143.44 GPa. Moreover, smaller diameter nanotubes tend to be stiffer and have higher Young's modulus but smaller critical strain values. The $Y_M$ Young's modulus for PG-yne monolayers is approximately 913 GPa at room temperature, while for PG-yneNTs, it can range from 497-789 GPa. 

\section{Acknowledgements}

This work received partial support from Brazilian agencies CAPES, CNPq, FAPDF, FAPESP, and FAPEPI. J.M.S and D.S.G thank the Center for Computational Engineering and Sciences at Unicamp for financial support through the FAPESP/CEPID Grant \#2013/08293-7. CENAPAD-SP (Centro Nacional de Alto Desenpenho em São Paulo - Universidade Estadual de Campinas - UNICAMP) provided computational support for L.A.R.J and J.M.S (proj634 and proj842). W.H.S.B. and J.M.S were supported by Laboratório de Simulação Computacional Cajuína (LSCC) at Universidade Federal do Piauí. L.A.R.J thanks the financial support from Brazilian Research Council FAP-DF grants $00193-00000857/2021-14$, $00193-00000853/2021-28$, and $00193-00000811/2021-97$, CNPq grants $302236/2018-0$ and $350176/2022-1$, and FAPDF-PRONEM grant $00193.00001247/2021-20$. L.A.R.J also thanks ABIN grant 08/2019 and Núcleo de Computação de Alto Desempenho (NACAD) for computational facilities through the Lobo Carneiro supercomputer. Additionally, Fundação de Apoio à Pesquisa (FUNAPE) provided financial support through Edital 02/2022 - Formulário de Inscrição N.4. This research also used computing resources and assistance from the John David Rogers Computing Center (CCJDR) in the Institute of Physics ``Gleb Wataghin'', at State University of Campinas.

\newpage
\bibliographystyle{elsarticle-num}
\bibliography{bibliografia.bib}
\end{document}